\documentclass[ams, prb, amssymb, nofootinbib, twocolumn, floatfix]{revtex4-1}
\usepackage{amsmath}
\usepackage{tabularx}
\usepackage{bm}
\usepackage{euscript}
\usepackage{graphicx}
\usepackage{color}
\usepackage{amsfonts}
\usepackage{exscale}
\usepackage{amsbsy}
\usepackage{subfigure}
\usepackage{textcomp}
\usepackage{comment}
\usepackage{slashed}
\usepackage{hyperref}

\newcommand{\beq}{\begin{equation}}
\newcommand{\eeq}{\end{equation}}
\newcommand{\be}{\begin{eqnarray}}
\newcommand{\ee}{\end{eqnarray}}

\begin{document}

\title{Two dimensional metallic phases from disordered QED$_3$}
\author{Pallab Goswami$^{1}$, Hart Goldman$^{2}$, and S. Raghu$^{3,4}$}
\affiliation{$^{1}$Condensed Matter Theory Center and Joint Quantum Institute, Department of Physics, University of Maryland, College Park, Maryland 20742- 4111 USA}
\affiliation{$^{2}$Department of Physics and Institute for Condensed Matter Theory, University of Illinois, Urbana-Champaign, Illinois 61801, USA}
\affiliation{$^{3}$Stanford Institute for Theoretical Physics, Stanford University, Stanford, California 94305, USA}
\affiliation{$^4$SLAC National Accelerator Laboratory, 2575 Sand Hill Road, Menlo Park, CA 94025, USA}
\date{\today}

\begin{abstract}

Metallic phases have been observed in several disordered two dimensional (2d) systems, including thin films near superconductor-insulator transitions and quantum Hall systems near plateau transitions. The existence of 2d metallic phases at zero temperature generally requires an  interplay of disorder and  interaction effects.  Consequently, experimental observations of 2d metallic behavior have largely defied explanation. We formulate a general stability criterion for strongly interacting, massless Dirac fermions against disorder, which describe metallic ground states with vanishing density of states. We show that (2+1)-dimensional quantum electrodynamics (QED$_3$) with a large, even number of fermion flavors remains metallic in the presence of weak scalar potential disorder due to the dynamic screening of disorder by gauge fluctuations. We also show that QED$_3$ with weak mass disorder exhibits a stable, dirty metallic phase in which both interactions and disorder play important roles. 
\end{abstract}

\maketitle


\section{Introduction}
All experimental systems simultaneously possess disorder and interactions. Nevertheless, the quantum dynamics of  disordered, interacting electron fluids remains poorly understood; our intuition is largely based on theories of  non-interacting, disordered fermions\cite{Wegner1976, Wegner1979, Abrahams1979,Lee1985,Zirnbauer1996}. The conclusion drawn from these studies is that metallic phases are largely absent\footnote{There are a few exceptions. Systems with time-reversal symmetry with strong spin-orbit coupling are immune to weak disorder. However, we will be neglecting spin-orbit effects in this work. } in two dimensions (2d) at zero temperature (${T}=0$). However, many experimental observations have suggested otherwise.  Metallic phases have been observed in a variety of 2d settings, including amorphous thin films near the superconductor-insulator transition\cite{Mason1999,Goldman2010}, two dimensional electron gases near quantum Hall transitions\cite{Wong1995,Jiang1989,Zudov2015}, and  certain 2d metal oxide semiconductor field effect transistors\cite{Kravchenko2003}. To the extent that these experimental observations reflect behavior at ${T}=0$, they imply that the 2d interacting, disordered electron fluid can exhibit non-trivial low energy phenomena.  

With very few exceptions\cite{Chakravarty1998,Plamadeala2014}, a clean metal possessing a finite density of states at the Fermi energy is highly unstable to disorder. As a result, the ballistic motion of electrons in the clean limit gives way to charge diffusion. For free fermions, the ultimate fate of such a diffusive metal at zero temperature depends on the dimensionality\cite{Abrahams1979} and the underlying discrete global symmetries\cite{Zirnbauer1996,Altland1997,Denis2002}. In the absence of interaction effects, diffusive behavior of a system with spin rotation symmetry ultimately gives way to localization in 2d.  Whether or not interaction effects alter this conclusion has remained controversial\cite{Finkelstein1984,Castellani1984,Castellani1998,Belitz1994,Miranda1997,PhysRevB.64.214204,Punnoose289,PhysRevB.89.235423}. 

In this article, we consider disorder and interaction effects in systems with vanishing density of states at the Fermi energy in the clean limit~\cite{Fradkin1986,Fradkin1986a}. We will focus on two dimensional, strongly interacting, massless Dirac fermions with zero carrier density. Despite the vanishing of the density of states in these systems, 
the dc conductivity at ${T}=0$ can be finite and {\it universal}. 
 Hence, we may rightly refer to such systems as `metals'. Although our primary motivation is to provide concrete examples of metallic phases in 2d, our analysis is relevant to a class of gapless spin liquids with algebraic or power-law correlations\cite{Rantner2001, Hermele2004,Hermele2005}.     
Below, we study the conditions under which such strongly interacting systems remain stable against disorder.  
Loosely speaking, the stability criteria we present here are analogs of the influential Harris criterion\cite{Harris1974} for critical points, applied to gapless states of strongly interacting fermions.

Using our stability criteria as an organizing framework, we provide concrete examples of various scenarios for resulting 2d metallic phases. These examples appear in the context of $N$ species of massless Dirac fermions coupled to a fluctuating $U(1)$ gauge field in two spatial dimensions (i.e. quantum electrodynamics in $2+1$D, denoted as QED$_3$), in the presence of quenched disorder. The clean theory is strongly coupled at long distances and lacks a quasiparticle description. 
However, we will consider the system either in a large $N$ expansion or in an $\epsilon$-expansion about 3 spatial dimensions\footnote{Note that an expansion about three spatial dimensions does not necessarily enable continuation to an arbitrarily small number of fermion species due to the possible onset of chiral symmetry breaking, whence the theory enters a strongly coupled, massive phase. Note also that we neglect non-perturbative effects which appear in the form of monopole operators by studying QED$_3$ with a non-compact gauge group.}, enabling us to control the effects of interactions. 
We show that in the presence of parity\footnote{reflection about $x$ or $y$ axis}, time-reversal, and chiral symmetries, QED$_3$ at large $N$ 
is stable against weak scalar potential disorder due to the dynamical screening of the disorder potential 
by the gauge field. 

In addition, we show that the same system perturbed by parity-preserving 
mass disorder (to be defined more precisely below) 
 retains metallic behavior, 
 but is governed  by a renormalization group (RG) fixed point with {\it finite} disorder and interaction strengths: the system is a dirty metal. While we restrict our attention to cases where parity  
and time-reversal symmetries are both preserved, 
the case in which these symmetries are broken  
is also of interest, as it is relevant to quantum Hall plateau transitions.  We shall consider such systems in a forthcoming publication.



\section{Stability criterion for incompressible metals}
The defining property of a metal is a finite dc conductivity at $T=0$.  The dc conductivity is defined by the following limit
\begin{equation}
\sigma_{dc} = \lim_{T \rightarrow 0} \lim_{\omega \rightarrow 0} \sigma(\omega, T)
\end{equation}
where $\sigma(\omega, T)$ is the ac conductivity at temperature $T$.  Note that the property above does not require a finite compressibility: even incompressible systems can exhibit a finite conductivity.  A useful example of such an `incompressible metal'   is clean graphene in the presence of Coulomb interactions.  At the neutrality point, the density of states vanishes.  Furthermore, momentum relaxation can occur in this system due to inelastic scattering from Coulomb interactions.  The inelastic scattering rate can be estimated from Fermi's golden rule:
\begin{equation}
1/\tau \sim \alpha^2 T,
\end{equation}
where $\alpha = e^2/v_F$ is the fine structure constant.  
The divergence of the lifetime is compensated by the vanishing of the density of states as $T \rightarrow 0$, and the Drude estimate for the dc conductivity is
\begin{equation}
\sigma_{dc} \sim  1/\alpha
\end{equation}
Since weak Coulomb interactions are marginally irrelevant, $\alpha$ vanishes slowly at low energies, and the system is actually a {\it perfect} conductor at zero temperature.  However, in the presence of disorder, the system acquires a finite density of states and crosses over into a conventional diffusive regime.  In other words, graphene is unstable to disorder.  By contrast, strong interaction effects are capable of forestalling such a crossover. In this paper, we will construct explicit examples of such behavior.  First, though, it will prove useful to determine a criterion for stability of such systems to weak disorder.  


We  begin with the Euclidean action $S = S_0 + S_{dis}$ in $d+1$ dimensional spacetime, where $S_0$ describes noninteracting  massless Dirac fermions, and $S_{dis}$ is a disorder perturbation.  
We may write $S_{dis}$ as
\begin{equation}
S_{dis} = \int d^d x d \tau \  V_{j}(x)  \bar \Psi_\alpha(x,\tau) \hat M_{\alpha\beta}^{j} \Psi_\beta(x,\tau),
\end{equation}
where $j$ labels the type of disorder, such as random chemical potential, mass, or static gauge fields (flux disorder). The variables $\{V_{j}\}$ are quenched random fields, i.e. they depend explicitly on space but not on time. They possess general power-law correlations specified by \begin{equation}
\label{eq: general disorder distribution}
\overline{V_{i} (x) V_{j}(x')}  \propto \delta_{ij} \frac{\Delta_j}{ \vert x - x' \vert^{ \chi_j}},
\end{equation}  
with $\chi_j = d$ corresponding to a Gaussian white noise distribution and where the bar denotes disorder averaging. 
Since the disorder field $\{V_{j}\}$ couples to a fermion bilinear (specified by the matrix $\hat M^{j}$), it has engineering mass dimension 1; as a consequence,  
 $[\Delta_j]=2-\chi_j$. 
Thus, $\Delta_j$ are dimensionless in any spatial dimension so long as $\chi_j=2$. In particular,  $\Delta_j$ are dimensionless in 2d when the disorder has Gaussian white noise correlations. It follows\cite{Goswami2016} that the system will be stable against the disorder perturbation of type $j$ provided that $\chi_j > 2$.  

The physics of an interacting theory,  however, 
can be 
richer.  Let $S_{int}$ be the action associated with the interactions such that $S= S_0 + S_{int} + S_{dis}$, and let us suppose that at low energies the interactions alter one of the $\chi_j$  (referred to as $\chi$ below) via 
\begin{equation}
\chi \rightarrow  \chi - 2 \eta \equiv \chi_{int}.
\end{equation}
Contributions to the deviation $2 \eta$ can occur either (i) when interactions screen the disorder\cite{Mirlin2008,Foster2008,Goswami2011}, or (ii) through fermion anomalous dimension effects. The latter effect can occur only if the disorder fields couple to {\it non-conserved} fermion bilinears. In either case, the criterion for the system to be stable against disorder at low energies is the altered condition  
\begin{equation}
\chi_{int} > 2.
\end{equation}
 In particular, when $\chi_{int} > \chi$, it is possible that the system with interactions becomes stable against weak disorder, while its non-interacting counterpart remains unstable\footnote{The converse is also possible when $\chi_{int} < \chi$, an example being mass disorder.}. The goal of the remainder of this article is to explore examples and consequences of these observations in the context of large $N$ QED$_3$.

\section{Large $N$ QED$_3$ with potential disorder}

We now consider a system of $N \gg 1$ flavors of massless four-component Dirac fermions coupled to an emergent $U(1)$ gauge field with both parity and time-reversal symmetries intact.  
Because we are using four-component fermions, there 
is 
an even number of Dirac nodes, as required by the doubling theorem, and each node in turn consists of $N$ two-component fermion flavors.  
The UV Euclidean action of such a system in the absence of disorder is
\begin{eqnarray}
S_0 = \int d^2 x d \tau \  \left[  \bar \Psi_j  \gamma_{\mu}  D_{\mu} \Psi_j + \frac{1}{4} f^2_{\mu \nu} \right]
\end{eqnarray} 
where $j=1, \cdots, N$ is the fermion flavor index, $\Psi^{\dagger} = \bar \Psi \gamma_0$, the covariant derivative is  $D_{\mu} = \partial_{\mu} + i ga_{\mu}$, $g$ is the gauge charge, and the field strength is
$f_{\mu \nu} = \partial_{\mu} a_{\nu} - \partial_{\nu} a_{\mu}$, and $\gamma_\mu$ are three mutually anticommuting, Hermitian Dirac matrices.  

We take the limit $N \rightarrow \infty$ in the standard fashion (see Appendix \ref{Appendix: QED3 review} for a brief review) by holding fixed the coupling $\alpha = g^2 N$. At leading order, the only diagrams that contribute involve fermion polarization bubbles, which dynamically screen the photon. 
As a result, the photon propagator takes the form $D(q) \sim 1/q$ at low energies.
Therefore, the infrared physics of QED$_3$ for sufficiently large $N$ is governed by a stable, conformal, interacting fixed point with $\alpha_\ast \sim 1$ and an $\mathcal O(1/N)$ fermion anomalous dimension which causes the quasiparticle description to break down. This fixed point theory is invariant under parity and time-reversal, as well as the full $U(2N)$ flavor symmetry, which is traditionally referred to as the chiral symmetry. We will refer to this fixed point theory as conformal QED$_3$, which supports  enhanced power law correlations for $U(2N)$ symmetry breaking Dirac mass terms. Intriguingly, this fixed point theory describes a strongly interacting metallic phase with a finite dc conductivity $\sigma \sim 1/\alpha_\ast$ for the flavor-singlet U(1) gauge current, and a finite dc conductivity $\sigma_f \sim 1/\alpha^2_\ast$ for the flavor currents. When the QED$_3$ theory describes a U(1) spin liquid, the flavor conductivity $\sigma_f$ corresponds to the spin conductivity.

\begin{figure}
\includegraphics[width=0.5\textwidth]{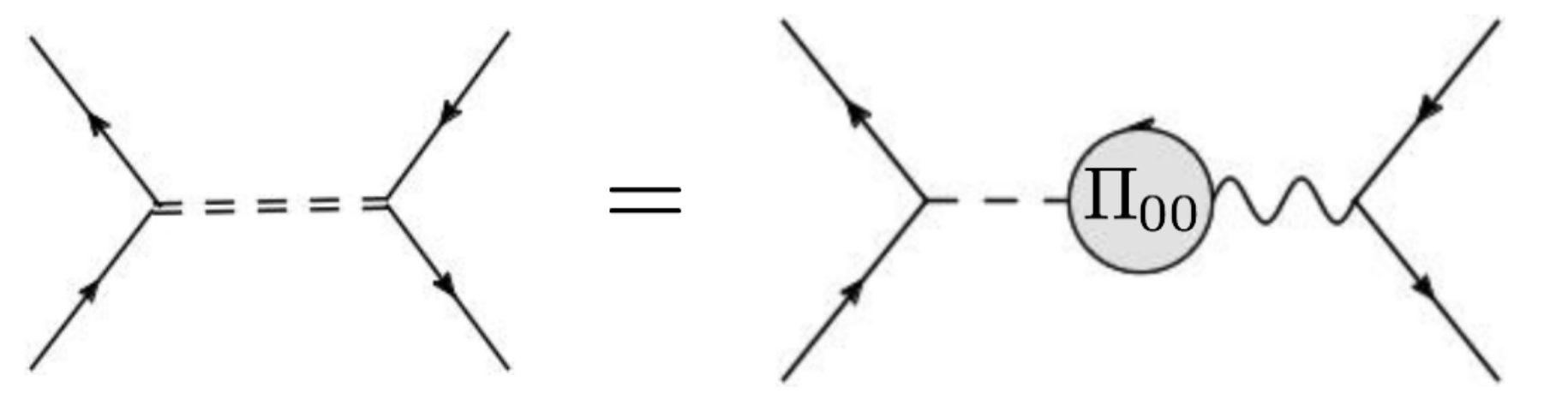}
\caption{The diagram which describes the screening of the random scalar potential by the fermion density-density polarization bubble at zero frequency. The dashed line represents the disorder interaction Eq. \eqref{eq: potential disorder four fermion interaction}. The shading of the bubble denotes a geometric series of bubble diagrams.}
\label{diagram_disorder}
\end{figure}

We now consider the effects of disorder. If we require that parity, time-reversal, and $U(2N)$ chiral symmetries are unbroken by the disorder, we may neglect mass and flux disorder and focus only on a random chemical or scalar potential, which couples to the local fermion density $\bar{\Psi}\gamma_0\Psi$. We take the disorder correlations to be Gaussian white noise ($\chi=d=2$): $
\overline{V(x)}= 0, \ \overline{ V(x) V(x')} = \Delta \delta^{(2)}(x-x').$
Disorder averaged quantities are obtained using the 
replica trick, where the disorder enters as a non-local interaction between $n_r$ distinct ``replicas" of the fermion fields:
\begin{equation}
\label{eq: potential disorder four fermion interaction}
S_{dis} = -\frac{\Delta}{2} \int d^2 x d \tau d \tau'  \bar{\Psi}^{a}_{i} \gamma_0 \Psi_{i}^a (x,\tau) \bar{\Psi}^{b}_j \gamma_0\Psi^b_j(x,\tau')
\end{equation}
where the superscripts $a,b = 1, \cdots, n_r$, and we take the replica limit $n_r \rightarrow 0$. Since this type of disorder describes fermion density fluctuations, it can be screened by the fermion density-density polarization bubble together with 
the longitudinal component of the photon propagator, as shown in Fig. \ref{diagram_disorder}. 
The RPA corrected longitudinal component $D_{00}$ of the photon propagator (with zero external frequency) screens the disorder but does {\it not} mix distinct replicas. Hence, such processes survive upon taking the replica limit. As shown explicitly in Appendix \ref{Appendix: potential disorder}, the resulting disorder correlation is altered by gauge fluctuations from Gaussian white-noise to a short-ranged form
\begin{equation}
\label{eq: screened correlator}
\overline{V(x) V(x')} \propto \frac{\Delta}{\vert x - x' \vert^3}.  
\end{equation} 
As a consequence of this screening, $\chi_{int} = 3$ (i.e. $2 \eta = -1$), and the clean conformal QED$_3$ fixed point  is stable against potential disorder.  It is our first example of a  2d metallic phase stabilized by interactions.  Such a system would have universal conductivity at $T=0$, which is determined by the conformal fixed point.  

It is interesting to contrast the conducting properties of graphene with $1/r$ interactions and the conformal QED$_3$ fixed point.  While both systems consist of fermions  with $1/r$ interactions, there are several key differences that are worth pointing out.  In the case of graphene, the fine structure constant vanishes in the long wavelength limit, whereas it reaches a fixed point value in the case of conformal  QED$_3$.  This difference leads to several consequences.  First, the QED$_3$ system will have a finite conductivity as opposed to graphene which will have a (logarithmically) diverging conductivity.  Additionally, as we have seen, the conformal QED$_3$ system is stable to weak potential disorder, whereas graphene is unstable.  These differences are reflected in the fact that graphene  exhibits fixed lines in the $\alpha-\Delta$ plane\cite{Foster2008}, whereas conformal QED$_3$ exhibits a fixed point.

\section{large $N$ QED$_3$ with Mass Disorder } 
We now present an example of a metallic phase 
with finite interactions and disorder correlations. Such a phase can be realized by considering a form of disorder to which a system of free Dirac fermions would be stable, but where interactions can produce a destabilizing effect. 
Our focus here will be on the simplest variety of such disorder, mass disorder of the form
\be S_{dis}=\int d^2xd\tau V(x)\bar{\Psi}_j(x,\tau)\Psi_j(x,\tau), \ee
with zero mean and Gaussian white noise correlations. 
A concrete realization of mass disorder in the context of graphene is a random potential that has opposite signs on each of the two sublattices of the honeycomb lattice. This perturbation preserves parity and time-reversal but breaks the sublattice, or spatial-inversion symmetry.
Since $\Psi$ is a four-component spinor, 
this term explicitly breaks the $U(2N)$ chiral symmetry of the clean theory down to $U(N)\times U(N)$, while respecting parity and time-reversal. Importantly, this type of disorder is unscreened by interactions, and at the tree level it is marginal at the conformal QED$_3$ fixed point.  

In the absence of interactions, mass disorder is marginally irrelevant, weakening at low energies.  
However, since the mass operator is not a conserved quantity, it can acquire an anomalous dimension due to interaction effects.  As discussed in Appendix \ref{Appendix: QED3 review}, the mass anomalous dimension at the conformal QED$_3$ fixed point is positive: $2\eta(\alpha_*)\sim \alpha_*/N$.  As a consequence, $\chi_{int} < \chi$, and, while the non-interacting system is stable to mass disorder, the interacting counterpart is {\it unstable}.    Thus,  we may expect a stable fixed point which retains {\it finite} disorder correlations as well as finite interactions. Similar behavior is expected for other types of random Dirac mass terms.

Since disorder breaks Lorentz invariance, the ratio $v/c$ of fermion to photon velocity is a running coupling along with disorder and interaction strengths.  We have found that  the simplest framework in which to obtain the flow of these couplings is to study the problem in $d=3-\epsilon$ space dimensions. Such a method can recover the physics of the conformal QED$_3$ fixed point in the clean limit for sufficiently large $N$, where $\alpha_\ast \sim \epsilon$ (see Appendix \ref{Appendix: QED3 review})\cite{DiPietroKomargodskiShamirStamou,Herbut2016CSB,GiombiKlebanovTarnopolsky,GRACEY1993415}. Since at tree-level $\left[ \alpha \right] = 3-d$ and $\left[ \Delta \right] = 2 - \chi$, we can fix $\chi = 2$ while varying the dimension of space. The problem of interest, Gaussian 
mass disorder, then corresponds to $\chi = 2, \epsilon=1$.  We will also define running couplings $\alpha, \Delta$ appropriate for $d=3-\epsilon$ space dimensions: $\bar \alpha =  \frac{g^2 N}{4 \pi^2 v}\Lambda^{-\epsilon}, \ \ \bar \Delta = \frac{\Delta}{2 \pi^2 v^2},$
where $\Lambda$ is a cutoff scale. The fact that these are the appropriate running couplings can be seen in perturbation theory. 

We obtain the RG flow equations by holding $c=1$ while allowing $v/c=v$  to run.  The action remains scale invariant after renormalization provided that we allow the dynamical critical exponent $z$ to run.  We study the equations to one-loop order in an epsilon expansion, including terms depending on $1/N$.     We discuss the details of this analysis in Appendix \ref{appendix: epsilon expansion} and simply quote the results here. At leading order in $\epsilon$, the flow equations of the running couplings are 
\begin{eqnarray}
z&=& 1+ \frac{1}{3}\bar \alpha \left(1-v^2 \right), \nonumber \\ 
\frac{d v}{d \ell} &=& v \left[ -\frac{2}{3}{ \bar \Delta} - \frac{ \bar \alpha}{ N} g_1(v) + \frac{1}{3}\bar \alpha \left( 1-v^2 \right) \right],  \nonumber \\ 
\frac{d \bar \alpha}{d \ell} &=& \bar \alpha \left[ \epsilon + \frac{2}{3} \bar \Delta - \frac{2}{3}\bar\alpha + \frac{\bar \alpha}{N} g_1(v) \right], \nonumber \\
 \frac{d \bar \Delta}{d \ell} &=& 2 \bar \Delta \left[ -\frac{4}{3} \bar \Delta + \frac{\bar \alpha}{N} g_2(v) \right],
\end{eqnarray}
where $g_1(v), g_2(v)$ are simple functions of $v$ and are provided in Appendix \ref{appendix: epsilon expansion}.    Fixed points are obtained by finding simultaneous zeroes of the above equations.  At {\it infinite} $N$, it can be seen by inspection that there is a nontrivial infrared-stable fixed point with $z=1, v=1, \bar\Delta = 0, \bar\alpha = 3 \epsilon/2$.  This corresponds to the clean conformal QED$_3$ system, and the system studied in the previous section would be obtained in the limit $\epsilon \rightarrow 1$.  However, when $N$ is large but {\it finite}, the clean QED$_3$ fixed point is destabilized in favor of an infrared-stable fixed point with finite $\bar\alpha, \bar\Delta$:
\begin{equation}
v_* \sim 1 - \frac{9}{8N}, \ \bar\alpha_*  \sim \frac{3\epsilon}{2}, \ \bar\Delta_* \sim \frac{27 \epsilon}{16N}, \ z_* \sim 1+ \frac{9 \epsilon}{8N}.
\end{equation}

There are several striking aspects of this solution. First, it represents a fixed point characterized by {\it finite interactions and finite disorder variance} which is stable to both disorder and interaction perturbations. Moreover, from $z_*$, we can determine the behavior of the density of states as a function of energy $\varepsilon$:
\begin{equation}
\rho(\varepsilon) \sim \varepsilon^{d/z_*-1} \sim_{\epsilon \rightarrow 1} \varepsilon^{1-\frac{9}{4N}}
\end{equation}
Using this, we deduce that the  compressibility vanishes as $\kappa \sim T^{1-9/4N}$, and the low temperature specific heat behaves as $c \sim T^{2-9/4N}$.  

The conductivity of this system likely remains finite and can be estimated as follows. The finite disorder variance here gives rise to elastic impurity scattering in addition to the inelastic scattering due to interactions. For a general dynamic scaling exponent $z$, $n \sim T^{d/z}$, $m \sim T^{(2/z-1)}$, and 
\begin{equation}
\frac{	1}{\tau} \sim (a_1 \Delta_* + a_2 \alpha^2_*) T ,
\end{equation}
where $a_1$ and $a_2$ are two constants of proportionality. Consequently, the conductivity behaves as $T^{(d-2)/z}$ and for finite mass disorder fixed point at $d=2$ we find  
\begin{equation}
\sigma = \frac{n \alpha }{m} \tau \sim \frac{\alpha_*}{(a_1\Delta_*+a_2 \alpha^2_*)}.
\end{equation}
This would imply that the system is in a \emph{dirty metallic phase}. In the future it will be interesting to improve on the above crude estimate by computing more quantitatively the conductivity at this fixed point.

It is also possible to estimate the diffusion constant $D$ of this system via the Einstein relation $\sigma\sim\alpha_*D(\varepsilon)\rho(\varepsilon)$. If we continue to $\epsilon=1$, $z_*<2$ for any $N>1$, so, as $T\rightarrow0$, $\rho(T)$ vanishes while $\sigma$ remains finite. Thus, from the Einstein relation, one would expect the diffusion constant to diverge at this dirty metallic fixed point, implying that this fixed point is characterized by so-called anomalous diffusion. In future work, we hope to understand this novel behavior more precisely.

Let us also comment on the dielectric properties at this fixed point.  
Both fermion and photon velocities display the momentum dependence $v(k) \sim k^{z_*-1}$ and $c(k) \sim k^{z_*-1}$, while their ratio $v/c$ becomes a constant less than one. The $k$ dependence of photon velocity originates from the following momentum dependent behaviors of effective permittivity and permeability: $\varepsilon \sim k^{-2\bar\alpha_*/3}$ and $\mu \sim k^{2 v^2 \bar\alpha_*/3c^2}$. 

To summarize, we have found, via an expansion about $3$ spatial dimensions, that the conformal phase of QED$_3$ is destabilized in the presence of mass disorder, giving way to a novel dirty metallic phase characterized by finite interaction strength and disorder correlations. We remind the reader that such a fixed point lacks a quasiparticle description due to the nonzero anomalous dimension of the Dirac fermions.

\section{Discussion}
In this article, we have considered the stability of strongly interacting metals with vanishing density of states against weak disorder. For concreteness, we have modeled the metal as two dimensional massless Dirac fermions coupled to an emergent $U(1)$ gauge field. In addition to establishing stable metallic phases due to interactions in 2d, our analysis is perhaps pertinent to a class of spin liquids with Dirac fermion spinon fields, also known as algebraic spin liquids, which may be described by the physics of conformal QED$_3$. To the extent that the large $N$ expansion is valid for such systems, based on our analysis, these systems ought to be stable against weak potential disorder. 
We have also shown that gauge fluctuations in conjunction with parity-preserving mass disorder lead to a dirty metal fixed point that can only occur in the presence of interactions and disorder. In the future, it will be interesting to study the gauge invariant spectral function, transport properties, and thermodynamic signatures of this dirty metal.  

While it is tempting to speculate that, for strong potential disorder, the systems considered here correspond to insulators, we stress that the true fate depends on the details of the microscopic problem and remains unknown. In this regime, the Dirac fermion obtains a {\it finite} density of states at zero energy. Hence, in this region, there is a finite scattering lifetime $\tau^{-1} \sim \bar \Delta \rho(0)$. The low energy behavior of such a system will be governed by a \emph{gauged, non-linear sigma model}. In the presence of a fluctuating gauge field, the true fate of  such nonlinear sigma model remains unclear and is worthy of further exploration. If this strong disorder phase indeed corresponds to an insulator, it may seem at first to contradict previous work,  
which argued that the absence of backscattering prevents localization\cite{Ryu2007}. However, it is important to note that these results are for theories of a single Dirac cone (or odd $N$); strong internode scattering can alter such conclusions.   

Lastly, let us speculate on the stability of 2d metals with a {\it finite} density of states in the clean limit.  In such systems, disorder effects typically overpower those due to intearactions at scales below $\tau^{-1} \sim k_F \Delta$.  However, the situation remains unclear in the strong coupling limit $\alpha \gg k_F$ when gauge fluctuations could dominate disorder effects. Indeed, 
it remains possible that by strongly screening disorder correlations, gauge fluctuations may also stabilize metals with Fermi surfaces.  We wish to study such possibilities more carefully in the future.  We also wish to consider the case of an odd number of two component Dirac fermions with a Chern-Simons gauge field.  It is conceivable that such an effective theory may provide a new description of quantum Hall plateau transitions, where interaction effects play an essential role.

{\it Note Added - }After the completion of our work, we learned about an interesting work by A. Thomson and S. Sachdev~\cite{Thomson2017}, which studies disorder effects in QED$_3$ with a different scope and methodology. We thank them for sharing a draft of their manuscript with us prior to publication. 
We have also recently become aware of Ref.~\onlinecite{Zhao2017}. Our disorder averaged diagrammatic perturbation theory and physical conclusions are different from those described there.

{\it Acknowledgments - } We thank S. Chakravarty, E. Fradkin, S. Kachru, M. Mulligan, C. Nayak, S. Sachdev, A. Thomson, and G. Torroba for discussions. P. G. is supported by JQI-NSF-PFC and LPS-MPO-CMTC. H. G. is supported by the National Science Foundation Graduate Research Fellowship Program under Grant No. DGE-1144245. S. R. is supported by the DOE Office of Basic Energy Sciences, contract DE-AC02-76SF00515.

\bibliography{metal}

\appendix
\onecolumngrid

\section{Review of Conformal QED$_3$}
\label{Appendix: QED3 review}
Recall that the action for QED$_3$ coupled to $N$ four-component Euclidean Dirac fermions is
\be
S=\int d^2xd\tau\left[\bar{\Psi}_j(\partial_\mu+iga_\mu)\gamma^\mu\Psi_j+\frac{1}{4}f^{\mu\nu}f_{\mu\nu}\right]
\ee
where $f_{\mu\nu}=\partial_\mu a_\nu-\partial_\nu a_\mu$. This theory is invariant under parity, charge-conjugation, and time reversal as well as an enhanced $U(2N)$ flavor symmetry historically referred to as chiral symmetry. The fact that the flavor symmetry is enhanced to $U(2N)$ can be seen in the following way. The above four-component spinor action can be related to the two-component spinor action by letting $\Psi_i=(\psi_i,\psi_{i+N})^T$, where $\{\psi_i\}$ is a set of $2N$ two-component spinors. The QED$_3$ action above is therefore invariant under flavor rotations of all of the two-component spinors $\psi_i$ amongst themselves, implying that the symmetry is $U(2N)$.

Under na\"ive scaling, the coupling $g$ is relevant and has mass dimension $[g]=1/2$. Thus, a controlled diagrammatic expansion in powers of the coupling is not possible directly in 2+1D. However, there are other perturbative methods one can appeal to in order to attempt to understand the low energy physics of this theory. One such approach is to study the theory for $N>>1$ and perform a controlled expansion in powers of $1/N$, holding $\alpha=Ng^2$ constant. Another approach is to perform an expansion in $\epsilon=3-d$, where $d$ is the number of spatial dimensions, and consider QED$_3$ as the limit $\epsilon\rightarrow1$. In this appendix, we review both of these approaches, which indicate that, for sufficiently large $N$, QED$_3$ flows to a stable, interacting conformal fixed point in the infrared (IR). 

First, consider the expansion in $1/N$. At leading order in $1/N$, the only diagrams are contributions to the photon self-energy involving fermion bubbles,
\be
\label{eq: fermion_bubble}
\Pi_{\mu\nu}=
\begin{gathered}
\includegraphics[width=0.15\textwidth]{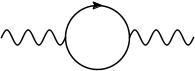}
\end{gathered}
\ee
To compute these diagrams, one can utilize the following Feynman rules for QED$_3$, which we will adopt for all of our large-$N$ calculations in these appendices. The bare fermion propagator is given by
\begin{equation}
G_0(k)=\frac{ik_\mu \gamma_\mu}{k^2},
\end{equation} 

To obtain the photon propagator, we add a gauge fixing term to the Lagrangian of the form
\be
\mathcal{L}_{\text{gauge fixing}}=-\frac{1}{2\xi}(\partial_\mu a^\mu)^2
\ee
For our purposes, it will be most convenient to choose Landau ($\xi=0$) gauge. In this gauge the photon propagator is
\be
D_{\mu\nu}(k)=\frac{\delta_{\mu\nu}-k_\mu k_\nu/k^2}{k^2}
\ee 
where we have suppressed flavor indices. Finally, the fermion-photon vertex is
\be
\label{eq: photon SE appendix A}
\begin{gathered}
\includegraphics[width=0.12\textwidth]{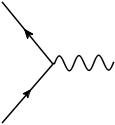}
\end{gathered}
=i\gamma^\mu\sqrt{\frac{\alpha}{N}}
\ee
Calculating the bubble diagram \eqref{eq: fermion_bubble}, one obtains a photon self-energy of the form
\be
\Pi_{\mu\nu}(k)=\Pi(k)\left(\delta_{\mu\nu}-\frac{k_\mu k_\nu}{k^2}\right)
\ee
where 
\be
\Pi(k)=\frac{\alpha}{8}k
\ee
Summing these bubbles into the photon propagator, one finds
\be
\label{eq: nonanalytic photon propagator}
D_{\mu\nu}(k)=\frac{\delta^{\mu\nu}-k^\mu k^\nu/k^2}{k^2+\Pi(k)}=\frac{\delta^{\mu\nu}-k^\mu k^\nu/k^2}{k^2+\alpha|k|/8}
\ee
This implies that $\alpha$ is renormalized by a factor of $\left(1+\frac{\Pi(k)}{k^2}\right)^{-1}$, which leads to a beta function for the dimensionless coupling $\tilde\alpha=\alpha/8|k|$ of the form
\be
\beta_{\tilde\alpha}=\tilde\alpha(1-\tilde\alpha)
\ee
which implies the existence of stable interacting IR fixed point at $\tilde\alpha_*=1$. This is the conformal QED$_3$ fixed point. One can also calculate sub-leading corrections in $1/N$, the most important consequence of this being that the fermion acquires a \emph{negative} anomalous dimension at the fixed point $\eta\propto-\tilde\alpha_*/N$. We note also that the parity-even fermion mass operator $\bar{\Psi}\Psi$ also has a \emph{negative} anomalous dimension. 

An expansion in $\epsilon=3-d$ will unambiguously yield the same result for $N$ greater than some critical value $N_{crit}$, below which the conformal QED$_3$ fixed point might become unstable due to the spontaneous breaking of chiral symmetry $U(2N)\rightarrow U(N)\times U(N)$. We will elaborate on this possibility further below. Within this $\epsilon$-expansion, one performs a loop expansion near $d=3$ and calculates RG equations to linear order in $\epsilon$. Such calculations are possible due to the fact that the mass dimension of $\alpha$ is now $\epsilon$. We may define a dimensionless running coupling $\bar\alpha=\Lambda^{-\epsilon}\frac{\alpha}{4\pi^2}$, where $\Lambda$ is an ultraviolet cutoff (note that this is a different running coupling from the one adopted in the large-$N$ case above). At one loop, the Ward identity implies that $\bar\alpha$ can only run due to the photon self-energy diagram \eqref{eq: photon SE appendix A}. Thus, from this diagram one immediately obtains a RG beta function for $\bar\alpha$
\be
\beta_{\bar\alpha}=-\frac{d\bar\alpha}{d\log\Lambda}=\epsilon\bar\alpha-\frac{2}{3}\bar\alpha^2
\ee
This equation implies that there exists a stable interacting IR fixed point at $\bar\alpha_*=\frac{3\epsilon}{2}$, i.e. we have recovered the conformal QED$_3$ fixed point. 

As mentioned above, an important and vexing question is what happens at small values of $N$. It is believed via, for example, analysis of Schwinger-Dyson equations at large-$N$ \cite{Pisarski1984,Appelquist1986,Appelquist1988,Kotikov2016,Kotikov2016a}, calculations using the $\epsilon$-expansion\cite{DiPietroKomargodskiShamirStamou,Herbut2016CSB}, arguments using entanglement monotonicity and the $F$-theorem\cite{GiombiKlebanovTarnopolsky,Grover2014}, as well as other approaches\cite{BraunGiesJanssenRoscher2014,FischerAlkoferDahmMaris2004,StrouthosKogut2009,KavehHerbut2005} that at a critical value of $N$ the conformal QED$_3$ fixed point is destabilized, and the theory flows to a strongly coupled, possibly massive fixed point. However, the conclusion that QED$_3$ is massive at small values of $N$ is far from proven. For example, a recent numerical study\cite{KarthikNarayanan2016} and the recently conjectured particle-vortex dualities for QED$_3$\cite{Aharony2016,Karch2016,Cheng2016} suggest that the conformal QED$_3$ fixed point or a CFT fixed point like it persists all the way down to a single flavor of four-component fermion (i.e. two flavors of two-component fermions). As mentioned in the Introduction, if this is true, or even if the conformal QED$_3$ fixed point remains stable at a small number of fermion flavors, then this would imply that our result for the stability of conformal QED$_3$ to Gaussian potential disorder takes on direct phenomenological relevance to the experimental search for algebraic spin liquids, as these systems have QED$_3$ (or even QCD$_3$) as an effective theory. 

\section{Bare Euclidean action and scaling conventions}
\label{Appendix: Notations}

While the previous Appendix directly discussed QED$_3$ in the clean limit, in this Appendix we will introduce our UV scaling conventions for QED in $d+1$ spacetime dimensions (QED$_{d+1}$) deformed by disorder. The bare Euclidean action of QED$_{d+1}$ with $N$ flavors of four-component fermions in the presence of a disordered fermion bilinear is
\begin{eqnarray}
\label{eq: replicated action}
S&=&S_0+S_{disorder}\\
S_0&=&\int d^{d}x d \tau \bigg [ \bar{\Psi}_j(\gamma_0D_0+v\vec{\gamma}\cdot\vec{D})\Psi_j +\frac{1}{4}f^{\mu\nu}f_{\mu\nu}\bigg] \\
S_{disorder}&=& \int d^dx d\tau V(x) \bar{\Psi}_j\hat{M}\Psi_j,
\end{eqnarray}
where $\tau$ represents imaginary time, $x_i$ are spatial variables, $D_\mu=\partial_\mu+ i\sqrt{\frac{\alpha}{N}}a_\mu$ is the covariant derivative, and $\alpha=Ng^2$, as defined in the previous Appendix.  
Chemical potential disorder corresponds to $\hat{M}=\gamma^0$ and mass disorder corresponds to $\hat{M}=1$. Since the disorder explicitly breaks Lorentz invariance, it is useful to introduce a dynamical critical exponent $z$ and define a fermion velocity $v$ (in units of $c=1$). 
In the absence of disorder, the engineering mass dimensions of the various quantities above are
\begin{eqnarray}
&&\left[ \vec{x} \right] =-1,   \left[ \tau  \right] = -z, \left[ \Psi \right] = d/2, \left[ a_0 \right] = \frac{\left(d+z-2 \right)}{2}, \nonumber \\
&& \left[ a_i \right] = \frac{\left(d-z\right)}{2}, \left[ \alpha \right] = z+2-d,  \left[ v \right] = 1-z.
\end{eqnarray}
The dimensions of the gauge fields are obtained by canonically normalizing with respect to electric component of the field strength.  While the assignment of dimensions does not affect the physics of the deep IR, the choice above has the appealing feature that the engineering dimension of the chemical potential is $\left[ \mu \right]  = z$, which is appropriate for an energy scale.  
Note that there is no Lorentz invariance when $v \neq 1$, even in the absence of disorder. 
 However, one can  easily show, as we do below, that the clean fixed points of QED (free or interacting) have $v=1$ and $z=1$, thereby exhibiting emergent 
 Lorentz invariance.

We will now move on to the scaling of the disorder. 
 The disorder potential is defined by its mean and variance, 
\begin{equation}
\label{eq: position space disorder propagator}
\overline{V(x)} = 0, \ \ \overline{V(x) V(x')} = \frac{\Delta}{|x-x'|^\chi},
\end{equation}
Note that we have assumed a general probabiity distribution for the disorder potential (which can be made precise using Riesz potential as done in Ref. \cite{Goswami2016}) with the Gaussian white noise distribution corresponding to $\chi=d$. 
Evidently, $\left[ V \right] = z$ since $V$ is an energy scale.   From the expression above for the disorder 2-point function, we see that 
\be
\left[ \Delta \right] = 2z-\chi.
\ee  

\section{Stability of the Conformal QED$_3$ Fixed Point to Potential Disorder}
\label{Appendix: potential disorder}

For computing disorder averaged correlation functions, we use the replica trick and write the replicated effective action as
\begin{eqnarray}
S=\sum_a^n\left[S_0[\Psi^a,a_\mu^a]+ \int d^2xd\tau V(x)\bar{\Psi}_{j}^a \gamma_0 \Psi_{j}^a + \frac{1}{2 \Delta} \int d^2x V(x) (-\nabla^2)^{y/2} V(x)\right] , \nonumber \\
\end{eqnarray}
where $a=1,2,..,n$ is the replica index and we will take $n \to 0$ at the end of the calculations and we have resumed focusing on $d=2$ spatial dimensions. We also have defined $y=d-\chi=2-\chi$ for convenience so that this action reproduces the disorder correlations \eqref{eq: position space disorder propagator}. 
We may treat the variables $V(x)$ as having a momentum-space propagator
\begin{equation}
\overline{V(\mathbf{k})  V(-\mathbf{k})} =  \frac{\Delta}{\vert \mathbf{k} \vert^y}.
\end{equation}
where, to emphasize the difference between spatial position (momentum) and time (frequency), we will bold the spatial position (momentum) vectors in this appendix. Integrating out the quenched degrees of freedom, we obtain a four-fermion interaction which is local space but nonlocal in imaginary time,  
\be
\label{eq: four fermion disorder interaction momentum space}
S_{dis}=-\frac{1}{2}\int \frac{d^2\mathbf{k}_1d^2\mathbf{k}_2d^2\mathbf{k}_3d\omega d\omega'}{(2\pi)^8}\bar{\Psi}^a_i(\mathbf{k}_1,\omega)\gamma_0\Psi^a_i(\mathbf{k}_2,\omega)\frac{\Delta}{|\mathbf{k}_1-\mathbf{k}_2|^y}\bar{\Psi}^b_j(\mathbf{k}_3,\omega')\gamma_0\Psi^b_j(\mathbf{k}_1-\mathbf{k}_2+\mathbf{k}_3,\omega')
\ee

In Appendix \ref{Appendix: QED3 review}, we discussed how the leading-order quantum corrections to QED$_3$ in the large-$N$ limit come from summing bubble diagrams. 
The effective, nonlocal four-fermion interaction Eq. \eqref{eq: four fermion disorder interaction momentum space} produced by the disorder is also screened via these bubbles at leading order in $1/N$. Of course, if we were to form bubbles from the disorder vertex alone, they would be proportional to the number of replicas $n$, which vanishes in the replica limit. However, if we form bubbles using only one disorder vertex and any number of $A_0$ vertices, as shown in Figure \ref{diagram_disorder}, then only flavor indices can be contracted. The full, screened four-fermion interaction can be thought of as being mediated by quenched degrees of freedom (which we will also call $V(\mathbf{x})$) with an effective propagator corrected by the factor $\left(1+ 2\frac{\Pi(\mathbf{k},0)}{\mathbf{k^2}} \right)^{-1}$, i.e. 
\begin{equation}
\overline{V(\mathbf{k})  V(-\mathbf{k})}  =   \frac{\Delta}{|\mathbf{k}|^{y}  \left(1+ 2\frac{\Pi(\mathbf{k},0)}{\mathbf{k^2}} \right)}=\frac{\Delta |\mathbf{k}|^{2-y}}{\mathbf{k}^2+2\Pi(\mathbf{k},0)}.
 \end{equation}
Since the bubble 
$\Pi(\mathbf{k},0)=\alpha |\mathbf{k}|/8$, 
$\Pi(\mathbf{k},0)$ is more relevant than the $\mathbf{k}^2$ term in the infrared limit, $k<<\alpha$. 
Consequently, in the IR limit, i.e. at the conformal QED$_3$ fixed point, 
the effective disorder propagator becomes
\begin{equation}
\overline{V(\mathbf{k})  V(-\mathbf{k})} = \frac{\Delta |\mathbf{k}|^{2-y}}{\mathbf{k}^2+\alpha |\mathbf{k}|/4} \rightarrow \frac{4\Delta}{ \alpha} \frac{1}{|\mathbf{k}|^{y-1}}.
\end{equation}
So, in position space, 
\be
\overline{V(\mathbf{x}')V(\mathbf{x})}\propto\frac{\Delta}{\alpha|\mathbf{x}-\mathbf{x}'|^{\chi+1}}
\ee
where we recall from above that $\chi=d-y$ and we recover Eq. \eqref{eq: screened correlator} for Gaussian disorder, i.e. $\chi=2$. Thus, in the deep IR, we can effectively replace the probability distribution  $\int d^dr V(\mathbf{r}) (-\nabla^2)^{y/2} V(\mathbf{r})/2\Delta$ by $\int d^dx V(x) |\nabla|^{(y+d-3)} V(x)/2 \Delta$. 
Since the conserved fermion density $\Psi^{\dagger}_j \Psi_j$ cannot acquire an anomalous dimension, 
the scaling dimension of $V$ remains unchanged: $\left[ V \right] = z$.  
Therefore, since $\alpha$ is marginal and $z=1$ at the conformal QED$_3$ fixed point, our expression for the screened disorder propagator implies
\begin{equation}
\left[ \Delta \right] = 2z + y - 3 = y-1
\end{equation}
Therefore, as long as $y<1$, disorder at the conformal QED$_3$ fixed point is irrelevant, indicating that the fixed point can overcome even stronger disorder than Gaussian (at the free fermion fixed point) white noise. 

\section{Disordered QED in $d=3-\epsilon$ Dimensions}
\label{appendix: epsilon expansion}
\subsection{Methodology}
While the large-$N$ approach adopted in the previous appendix is a powerful tool to study the stability of the conformal QED$_3$ fixed point to potential disorder, it can become computationally cumbersome in situations where one is interested in studying subleading effects in $1/N$. This is because disorder will lead to the running of the fermion velocity $v$ (or, alternatively, the speed of light $c$), and subleading calculations in $1/N$ must be performed using the nonanalytic photon propagator \eqref{eq: nonanalytic photon propagator}. Thus, the $v$ $(c)$-dependent form of the fermion propagator at $\mathcal{O}(1/N)$ is complicated, making further study of the perturbative effects of disorder needlessly difficult. Instead, we will now study QED$_d$ in $d=3-\epsilon$ spatial dimensions, where we hold fixed $\chi=2$, i.e. the disorder correlations will be fixed as
\be
\label{eq: disorder correlations in d=3-epsilon}
\overline{V(\mathbf{x}')V(\mathbf{x})}\propto\frac{\Delta}{|\mathbf{x}-\mathbf{x}'|^{2}}
\ee
This implies that, for any $\epsilon$, disorder will be marginal at a Lorentz invariant ($z=1$) fixed point. Thus, at $d=2$ ($\epsilon=1$), these correlations will correspond to Gaussian disorder. Note that we can view this procedure as a special case of a double-$\epsilon$ expansion in which there is an additional $\epsilon'=2-\chi$ where the case of interest corresponds to $\epsilon'=0$. 

Below, we calculate the RG equations governing the behavior of $v$, $\alpha$, and $\Delta$ to linear order in $\epsilon$ for both parity-preserving mass and potential disorder. In the former case, we find an infrared stable, disorder and interaction controlled fixed point. In the latter case, we will corroborate the conclusions of Appendix \ref{Appendix: potential disorder} and argue that there exists a metal-insulator crossover transition at sufficiently large values of the disorder variance.

\subsection{Feynman Rules} 
\label{Appendix: Feynman rules}
In what follows, we will choose Feynman ($\xi=1$) gauge for the photon propagator,  
leading to the following Feynman rules associated with the replicated action \eqref{eq: replicated action}. 
\be
\begin{gathered}
\includegraphics[width=0.12\textwidth]{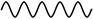}
\end{gathered}
&=&D_{\mu\nu}(p)=\frac{\delta^{\mu\nu}}{k^2} \\
\begin{gathered}
\includegraphics[width=0.12\textwidth]{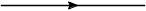}
\end{gathered}
&=&G_0(p)=\frac{i\slashed{k}}{k^2}\\
\begin{gathered}
\includegraphics[width=0.12\textwidth]{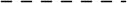}
\end{gathered}
&=&\frac{\Delta}{|\mathbf{k}|^{1-\epsilon}} \\
\label{eq: QED3_vertex}
\begin{gathered}
\includegraphics[width=0.12\textwidth]{Photon_Vertex.jpg}
\end{gathered}
&=&i\gamma^\mu\sqrt{\frac{\alpha}{N}}\\
\begin{gathered}
\includegraphics[width=0.12\textwidth]{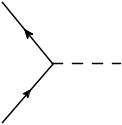}
\end{gathered}
&=&-\hat{M}
\ee
where we have suppressed flavor and replica indices. Again note that the photon in the QED3 vertex \eqref{eq: QED3_vertex} can only couple to fermion fields carrying its replica index, while the quenched degrees of freedom (which don't have replica indices themselves) can couple to fermions carrying any replica index. 


\subsection{Renormalization Group Calculation for Parity-Preserving Mass Disorder} 

We will compute the running of the disorder variance $\Delta$, the fine structure constant $\alpha=N_fg^2$, and the fermion velocity $v$ at one loop order utilizing a smooth spatial momentum cutoff. 
We remind the reader that we consider only diagrams which are non-vanishing in the replica limit. For simplicity, we will suppress replica and flavor indices as well as terms beyond the leading logarithmic divergences. Using the Feynman rules of Appendix \ref{Appendix: Feynman rules} and introducing the dimensionless running couplings $\bar{\Delta}=\frac{\Delta}{2\pi^2v^2}$ and $\bar{\alpha}=\frac{\alpha}{4\pi^2v}\Lambda^{-\epsilon}$, we begin with photon self-energy
\be
\begin{gathered}
\label{eq: bubble}
\includegraphics[width=0.23\textwidth]{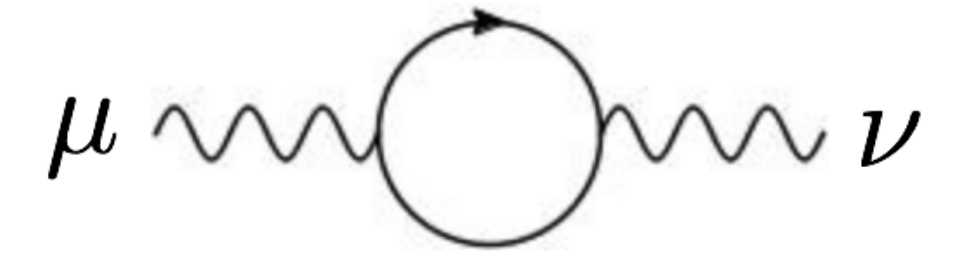}
\end{gathered}
&=&\frac{2 \bar{\alpha}}{3} d\ell [(k^2\delta^{\mu\nu}-k^\mu k^\nu)+(v^2-1)(1-\delta^{\mu 0})(1-\delta^{\nu 0})(\mathbf{k}^2\delta^{\mu \nu}-k^\mu k^\nu)]
\ee
where $d\ell=-d\log\Lambda$. Note that repeated indices are \emph{not} summed over in this expression. The contributions to the fermion self-energy are
\be
\begin{gathered}
\label{eq: fermion SE}
\includegraphics[width=0.15\textwidth]{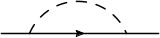}
\end{gathered}
&=&i\bar\Delta d\ell\left(\gamma^0\omega+\frac{1}{3}v\vec{k}\cdot\vec{\gamma}\right)\\
\label{eq: fermion SE 2}
\begin{gathered}
\includegraphics[width=0.15\textwidth]{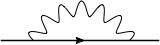}
\end{gathered}
&=&+i\frac{\bar{\alpha}}{N} d\ell\left(f_\omega(v)\omega\gamma^0+f_k(v)\vec{k}\cdot\vec{\gamma}\right) 
\ee
where we define $f_\omega(v)=\frac{(3v^2-1)v}{(1+v)^2}$ and $f_k(v)=\frac{(2+v)(1+v^2)}{3(1+v)^2}$. 

Gauge invariance requires that the corrections to the electron-photon vertex must match the fermion self-energy, i.e. $\frac{d\Sigma}{dp_\mu}=\Gamma_{\bar{\psi}\slashed{A}\psi}$, where $\Sigma$ is the fermion self-energy and $\Gamma_{\bar{\psi}\slashed{A}\psi}$ is the electron-photon vertex function. Explicitly,
\be
\begin{gathered}
\label{eq: photon vertex corrections}
\includegraphics[width=0.2\textwidth]{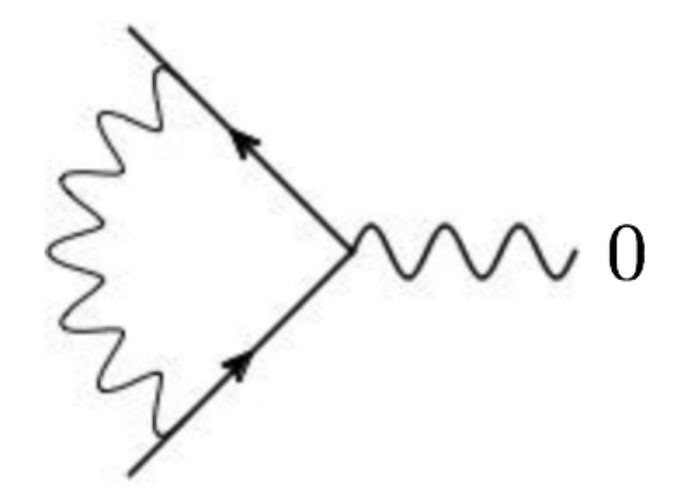}
\end{gathered}
=i\sqrt{\frac{\alpha}{N}}\gamma^0\frac{\bar{\alpha}}{N}f_\omega(v) d\ell&,&
\begin{gathered}
\includegraphics[width=0.2\textwidth]{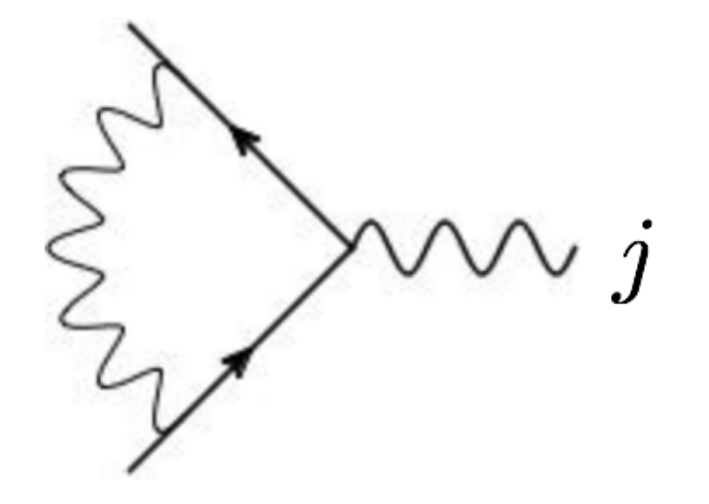}
\end{gathered}
=i\sqrt{\frac{\alpha}{N}}\gamma^j\frac{\bar{\alpha}}{N}f_k(v) d\ell\\
\begin{gathered}
\includegraphics[width=0.2\textwidth]{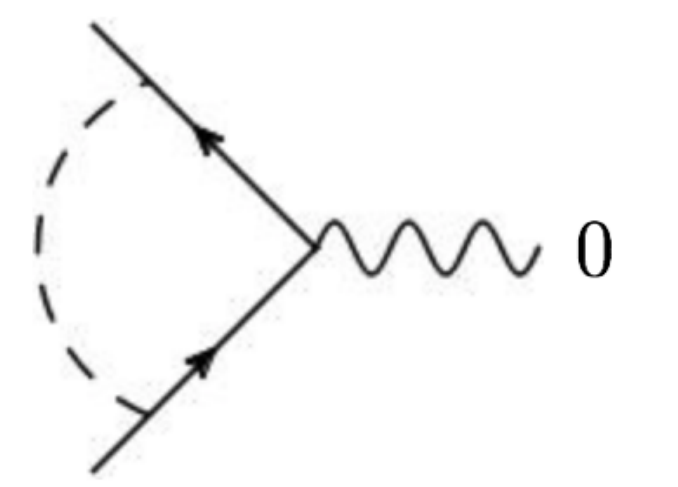}
\end{gathered}
=i\sqrt{\frac{\alpha}{N}}\gamma^0\bar\Delta d\ell&,&
\begin{gathered}
\includegraphics[width=0.19\textwidth]{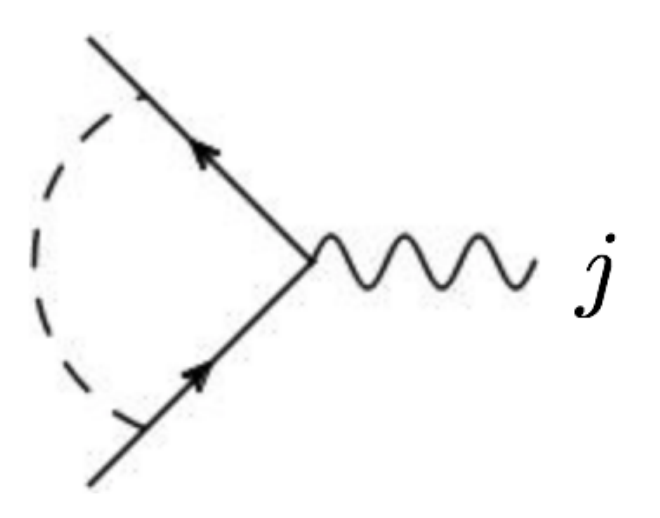}
\end{gathered}
=i\sqrt{\frac{\alpha}{N_f}}\gamma^j\frac{1}{3}\bar\Delta d\ell
\ee
where $j=1,...,d$. Finally, The corrections to the disorder vertex are
\be
\label{eq: disorder vertex corrections}
\begin{gathered}
\includegraphics[width=0.15\textwidth]{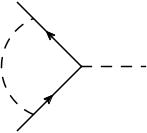}
\end{gathered}
=\bar\Delta d\ell
&,&
\begin{gathered}
\includegraphics[width=0.15\textwidth]{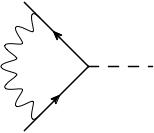}
\end{gathered}
=-f_\Delta(v)\frac{\bar{\alpha}}{N} d\ell
\ee
where $f_\Delta(v)=\frac{3v^2+1}{1+v}$. Note that these are the only vertex corrections contributing to the renormalization of the four-fermion disorder interaction, as the ladder and crossed diagrams associated with this interaction cancel at one loop order. 

We can now bring these corrections together to calculate the running of $\bar\Delta$, $\bar\alpha$, and $v$. The photon self-energy \eqref{eq: bubble} renormalizes the electric and magnetic field strengths differently, it is necessary to define a dynamical scaling exponent $z$. Using the na{\"\i}ve scaling conventions of Appendix \ref{Appendix: Notations} with $\chi=2$, we find that $z$ must be
\be
z=1+\frac{1}{3}\bar\alpha(1-v^2)
\ee
Using this expression for $z$ and the fermion self-energy diagrams \eqref{eq: fermion SE}-\eqref{eq: fermion SE 2}, we can immediately deduce the RG beta function for $v$
\be
\beta_{v}&=&\frac{d v}{ d\ell}=v\left[(z-1)-\frac{\bar\alpha}{N}g_1(v)-\frac{2}{3}\bar\Delta\right]\nonumber\\
&=&v\left[\frac{1}{3}\bar\alpha(1-v^2)-\frac{\bar\alpha}{N}g_1(v)-\frac{2}{3}\bar\Delta\right]
\ee
where we define $g_1(v)=f_\omega(v)-f_k(v)$. Using this beta function, we can obtain expressions for the beta functions for the dimensionless running couplings $\bar\Delta$ and $\bar\alpha$ using the vertex corrections \eqref{eq: photon vertex corrections}-\eqref{eq: disorder vertex corrections},
\be
\beta_{\bar\Delta}&=&\frac{d\bar\Delta}{ d\ell}=2\bar\Delta\left[-\frac{4}{3}\bar\Delta+\frac{\bar\alpha}{N}g_2(v)\right]\\
\beta_{\bar\alpha}&=&\frac{d\bar\alpha}{ d\ell}=\bar{\alpha}\left[\epsilon+\frac{2}{3}\bar\Delta-\frac{2}{3}\bar\alpha+\frac{\bar\alpha}{N}g_1(v)\right]
\ee
where we have defined $g_2(v)=f_\Delta(v)-f_k(v)$. These expressions imply that $d=3-\epsilon$ dimensional QED in the presence of parity-preserving mass disorder runs to a stable, finite disorder fixed point at 
\begin{equation}
v_* \sim 1 - \frac{9}{8N}, \ \bar\alpha_*  \sim \frac{3\epsilon}{2}, \ \bar\Delta_* \sim \frac{27 \epsilon}{16N}, \ z_* \sim 1+ \frac{9 \epsilon}{8N}
\end{equation}
This is the only nontrivial fixed point of the theory with $0<v<1$, and it corresponds to a dirty metallic phase. Notice that this fixed point becomes the clean conformal QED$_3$ fixed point in the limit $N\rightarrow\infty$.

\subsection{Renormalization Group Calculation for Potential Disorder}

We now comment on the case of potential disorder in $d=3-\epsilon$ spatial dimensions with correlations given by \eqref{eq: disorder correlations in d=3-epsilon}. There are two essential differences between this case and that of mass disorder. The first difference is in the corrections to the fermion self-energy and the disorder vertex in which disorder runs in the loop
\be
\begin{gathered}
\includegraphics[width=0.15\textwidth]{Fermion_SE_Disorder.jpg}
\end{gathered}
&=&i\bar\Delta d\ell(\gamma^0\omega-\frac{1}{3}v\vec{k}\cdot\vec{\gamma})\\
\begin{gathered}
\includegraphics[width=0.15\textwidth]{Disorder_VertexCorrection_Disorder.jpg}
\end{gathered}
&=&-\bar\Delta\gamma^0 d\ell
\ee
The sign differences in these diagrams imply that the anomalous dimension of the density operator $\bar\Psi\gamma^0\Psi$ vanishes, as required by the Ward identity. 

The second major difference between potential and mass disorder case is that potential disorder is screened by interactions via the diagram
\be
\begin{gathered}
\includegraphics[width=0.25\textwidth]{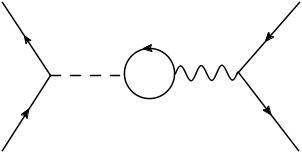}
\end{gathered}
=-\frac{2}{3}\bar\alpha\gamma^0_{\alpha\beta}\gamma^0_{\delta\epsilon}\Delta d\ell
\ee
where the Greek indices are the spinor indices of the external fermion lines. 

Thus, we obtain a new set of beta functions
\be
\beta_{v}&=&v\left[\frac{1}{3}\bar\alpha(1-v^2)-\frac{\bar\alpha}{N}g_1(v)-\frac{4}{3}\bar\Delta\right]\\
\beta_{\bar\alpha}&=&\bar\alpha\left[\epsilon+\frac{4}{3}\bar\Delta-\frac{2}{3}\bar\alpha+\frac{\bar\alpha}{N}g_1(v)\right] \\
\beta_{\bar\Delta}&=&2\bar\Delta\left[\frac{4}{3}\bar\Delta-\frac{2}{3}\bar\alpha-\frac{\bar\alpha}{N}g_1(v)\right]
\ee
Here the only nontrivial fixed point is the clean conformal QED$_3$ fixed point
\begin{equation}
v_*=1, \ \bar\alpha_*=\frac{3\epsilon}{2}, \ \bar\Delta_*=0, \ z_*=1
\end{equation}
Importantly, this fixed point is infrared-stable, implying that conformal QED$_3$ is stable to potential disorder. Recall that we also obtained this result in the large-$N$ analysis of Appendix \ref{Appendix: potential disorder}. These beta functions also indicate that, for sufficiently large disorder strength $\bar\Delta$, there will be a  crossover, beyond which the disorder variance runs strong, and our perturbative methods break down.

\end{document}